**Complejidad descriptiva y computacional en máquinas de Turing pequeñas**

Joost J. Joosten[1], Fernando Soler-Toscano[1], Hector Zenil[2,3]

[1] Universidad de Sevilla. Departamento de Filosofía y Lógica.

{jjoosten,fsoler}@us.es

[2] Laboratoire d'Informatique Fondamentale de Lille (CNRS), Université de Lille I

[3] Wolfram Research, Inc.

hectorz@wolfram.com

## Resumen

En este trabajo ofrecemos algunos de los resultados obtenidos al estudiar las funciones calculadas por máquinas de Turing con un alfabeto de dos símbolos y un número de estados entre 1 y 3. Presentamos detenidamente el modelo de máquina de Turing empleado, donde la cinta es ilimitada en un solo sentido, y no existe estado de parada. Mostramos gráficamente cómo es posible representar la función calculada por una máquina, así como las computaciones realizadas para ello. Vemos que al aumentar el número de estados, si bien es posible que algunas funciones se calculen con mayor rapidez, por lo general aumenta considerablemente el tiempo medio de computación.

## Introducción

Las máquinas de Turing constituyen tal vez el ejemplo más conocido de dispositivo de computación universal, lo que según la tesis de Church-Turing significa que para cualquier función efectivamente calculable existe una máquina de Turing que la calcula. Especialmente interesantes son las máquinas de Turing universales, capaces de simular la computación de cualquier otra máquina de Turing. Ahora bien, ¿qué funciones pueden calcular las máquinas de Turing que sabemos (o creemos) no universales, debido a la simplicidad de su diseño (pequeño número de estados)? Estas son las máquinas de Turing pequeñas, objeto de nuestro estudio.

Nuestro trabajo parte de ciertas conjeturas teóricas pero sigue una metodología empírica. Hemos programado un simulador que ha ejecutado grandes conjuntos de máquinas de Turing, produciendo gran cantidad de datos. Ofrecemos aquí algunas de las conclusiones que hasta el momento hemos obtenido del análisis de estos datos. Nos interesa especialmente estudiar cuál es la complejidad de las funciones que calculan máquinas tan simples y cómo aumenta o disminuye el coste en tiempo cuando agregamos más recursos (estados) a las máquinas.

**Preliminares**

Sean $\Sigma_1$ y $\Sigma_2$ dos conjuntos de símbolos a los que llamamos *alfabetos*. Mediante $\Sigma_i^*$ representamos el conjunto de todas las cadenas de símbolos de longitud finita del alfabeto $\Sigma_i$ (incluyendo la cadena vacía). Un *problema* es una relación entre cadenas de (un subconjunto de) $\Sigma_1^*$ (entradas) y (un subconjunto de) $\Sigma_2^*$ (salidas). Por ejemplo, la suma de números naturales se puede entender como el problema que asigna a cada cadena de la forma $\{0,\ldots,9\}^* + \{0,\ldots,9\}^*$ una única cadena de $\{0,\ldots,9\}^*$ de acuerdo con los axiomas de la adición de números enteros. En este caso la relación entre $\Sigma_1^*$ y $\Sigma_2^*$ es una función, pero no tiene por qué ser así en todos los casos, ya que hay problemas que admiten solución múltiple.

En función de cómo sean los conjuntos de cadenas de entrada y salida y la asignación entre ambos, podemos distinguir distintos tipos de problemas. Los más importantes:

- Problemas de decisión. En este caso la solución del problema será "si" o "no" (dos cadenas de $\Sigma_2^*$ con este significado). El problema queda determinado por el conjunto de cadenas de $\Sigma_1^*$ cuya solución es "si". Un ejemplo es el problema SAT en lógica proposicional, donde se determina si una fórmula es satisfacible.
- Problemas de búsqueda. La solución es un elemento de $\Sigma_2^*$ que cumple cierta relación con el elemento de $\Sigma_1^*$ que se toma como entrada. El ejemplo que dimos de la suma es un problema de este tipo.

Decimos que un problema es *computable* cuando existe un procedimiento efectivo (*algoritmo*) que permite obtener, para cualquier entrada de $\Sigma_1^*$ una cadena (o la única) de $\Sigma_2^*$ que le corresponde como *solución* del problema.

Un *modelo de computación* es un conjunto de estructuras de datos sobre los que están definidas ciertas operaciones que permiten implementar la noción del algoritmo. Entre los modelos de computación clásicos están las máquinas de Turing, el cálculo lambda y las funciones recursivas. Todos ellos tienen el mismo poder en cuanto a los problemas que pueden resolver. La tesis de Church-Turing afirma que esta capacidad computacional coincide con la noción intuitiva de resolubilidad de un problema. En este sentido se toman estos modelos como *universales*. Los problemas que no pueden ser resueltos en estos modelos se llaman *indecidibles*, son problemas para los que no existe un algoritmo que resuelva todas las instancias. En el caso de los problemas de decisión, son *semi-decidibles* aquellos para los que existe un algoritmo que resuelve las instancias positivas pero no las negativas, o viceversa.

Los problemas computables se pueden clasificar de diversas formas en función de su *complejidad* (cantidad necesaria de recursos para resolverlos). Existen distintas nociones de complejidad:

- Complejidad computacional. Atiende a la cantidad de tiempo o espacio necesaria para resolver el problema. Generalmente espacio y tiempo se definen en función del tamaño de la entrada (cadena de $\Sigma_1^*$ ) y según sea esta función hablamos de complejidad constante, lineal, polinomial, exponencial, etc., en el tiempo o en el espacio.
- Complejidad descriptiva. Hace referencia al tamaño mínimo del programa (algoritmo) que resuelve el problema. Cuando introduzcamos las máquinas de Turing, el tamaño se medirá por el número de símbolos y estados, ya que éstos determinarán el tamaño de la tabla de transiciones.

En este trabajo vamos a presentar algunos detalles de la investigación que estamos realizando en la relación entre complejidad computacional y complejidad descriptiva en máquinas de Turing pequeñas. Primero, mostraremos algunos detalles del formalismo que utilizaremos.

**Máquinas de Turing**

Las máquinas de Turing son el modelo de computación más conocido, debido a que es un modelo que tiene una representación física cuya motivación fue la descripción de un humano que calculara con lápiz y papel, profesión que se conocía como calculador o computador. Podemos verlas como una abstracción de nuestras computadoras. Disponen de una cinta de longitud ilimitada dividida en celdas discretas (análoga a la tira de papel donde escribía el computador humano) sobre la que se sitúa una cabeza capaz de leer y escribir en la celda donde se encuentra. La máquina sólo lee y escribe un conjunto finito de símbolos conocido como su *alfabeto*. Entre estos símbolos hay uno llamado usualmente *blanco* que es el que por defecto llena todas las celdas de la cinta. Existe un conjunto finito de estados en los que puede encontrarse la máquina. Uno de tales estados es el *estado inicial* desde el que comienzan todas las computaciones. También suele haber un estado de parada, que cuando se alcanza se termina la computación. En cada paso de computación, la máquina de Turing:

1. Lee el símbolo escrito en la celda sobre la que se encuentra la cabeza.
2. En función del símbolo leído y del estado actual de la máquina:
    - Escribe un nuevo símbolo en la celda (puede ser igual al que había).
    - Se desplaza una posición a la izquierda o derecha sobre la cinta.
    - Cambia de estado (o permanece en el mismo).

Así se continúa hasta llegar al estado de parada. Lo que caracteriza las computaciones de una máquina de Turing es su tabla de transiciones. Si vemos la enumeración anterior, el comportamiento en cada paso de computación dependerá del estado en que se encuentra la máquina de Turing y el símbolo leído. Por tanto, la tabla de transiciones tiene tantas entradas como el producto del número de estados por el número de símbolos que la cabeza lee y escribe. La ilustración 1 muestra la tabla de transiciones de una máquina de Turing con dos estados {1, 2} (representados por una aguja apuntando hacia arriba y hacia abajo, respectivamente) y dos símbolos, blanco y negro. El estado inicial es el número 1.

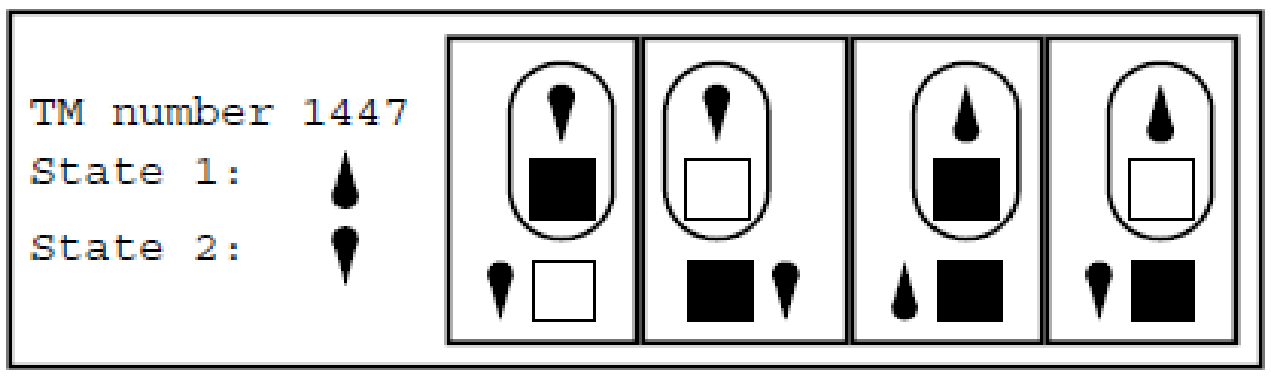

*Ilustración 1: Tabla de transiciones de una máquina de Turing*

Esta máquina tiene asignado el número 1447, ya que es posible enumerar las máquinas de Turing. El lector que quiera obtener más información sobre la enumeración que usamos en este trabajo y otras posibles puede acudir a Wolfram (2002) o Joosten (2010).

Se puede observar que por cada par (estado, símbolo) hay definida una transición. Cada una de las cuatro celdas representa una entrada en la tabla de transiciones. Los símbolos que están dentro del óvalo indican el par (estado, símbolo) para el que se define la transición. En la parte inferior se indica el nuevo símbolo que se escribe (color del cuadrado), nuevo estado (orientación de la aguja) y desplazamiento de la cabeza (aguja a la izquierda o derecha del cuadrado).

Veamos cómo se leen estas transiciones. Si vemos las dos de más a la derecha, nos dicen que cuando la máquina está en el estado 1 sobre una celda blanca: escribe negro en la celda, se desplaza a la izquierda y pasa al estado 2. Cuando la máquina está en el estado 1 sobre una celda negra: escribe negro, se mueve a la izquierda y permanece en el estado 1.

En el modelo de máquinas de Turing que empleamos no existe estado de parada y la cinta tiene una posición inicial, sobre la que originalmente se sitúa la cabeza, siendo ilimitada sólo hacia la izquierda. A la derecha de la posición inicial no hay ninguna celda. Al no haber estado de parada, necesitamos un criterio de terminación de las computaciones. Nosotros consideramos que una computación termina cuando la cabeza se encuentra en la posición inicial e intenta moverse hacia la derecha. En esta situación, donde la cabeza "escaparía" de la cinta, consideramos que termina la computación.

Podemos representar gráficamente la evolución de la cinta en cada paso de computación. Eso es lo que muestra la siguiente ilustración:

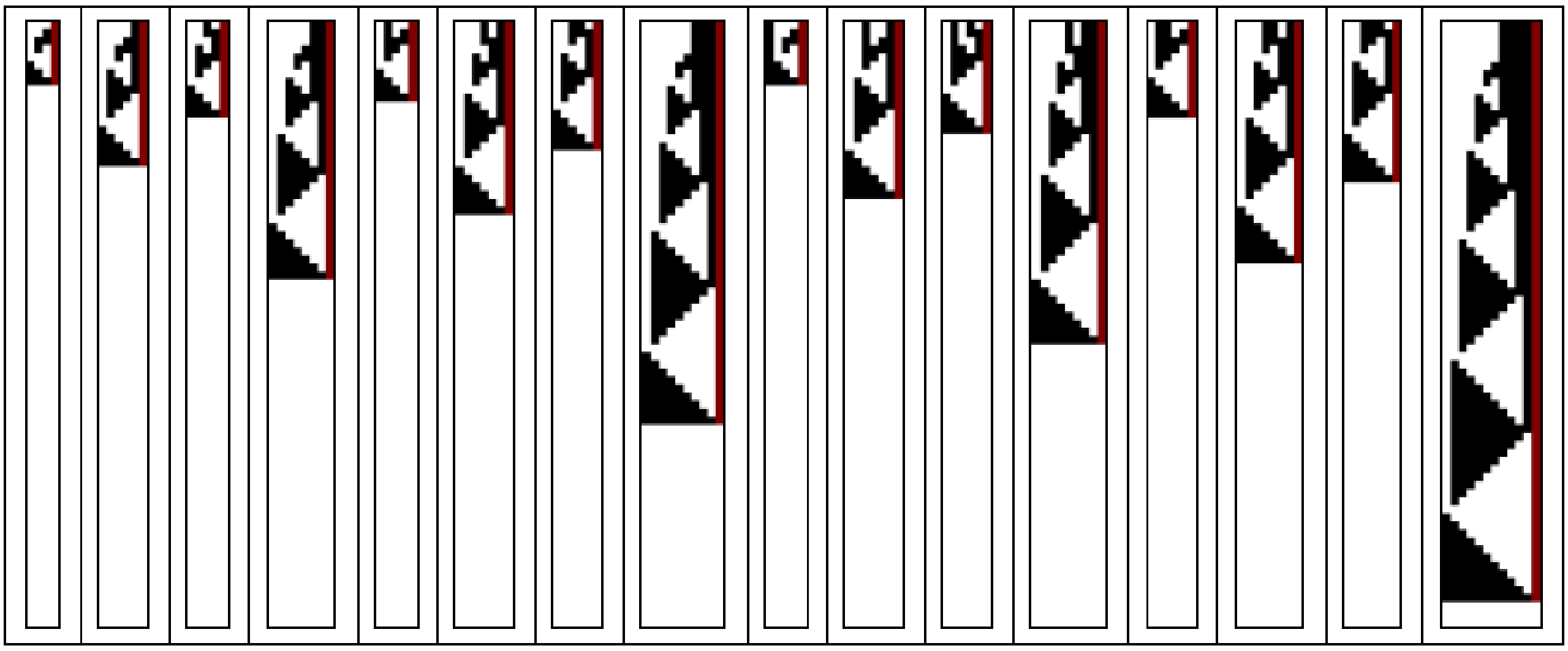

*Ilustración 2: Computaciones de la máquina 1447*

Aparecen representadas las computaciones que realiza la máquina de Turing número 1447, cuya tabla de transiciones mostramos más arriba, cuando la cinta contiene, al inicio, la representación binaria de los números entre 0 (la de más a la izquierda) y 15 (la de más a la derecha). Formalmente, podemos definir la computación que realiza una máquina de Turing como la secuencia de entradas a la tabla de transiciones que aplica hasta detenerse. Esto incluye implícitamente los símbolos que lee y escribe, los estados por los que pasa y los movimientos de la cabeza. Obviamente, esta secuencia de transiciones depende tanto de la (parte visitada de la) cinta de entrada como de la propia tabla de transiciones. Pese a que la cinta es unidimensional, al representar su evolución en el tiempo (cada una de las líneas representa el estado de la cinta en cada paso de computación) obtenemos los 16 diagramas bidimensionales que vemos en la figura. La celda de más a la derecha, que siempre aparece coloreada, representa el final de la cinta. La cabeza se sitúa originalmente en la segunda posición al contar desde la derecha (primera posición de la cinta). Como dijimos, la cinta es ilimitada, aunque sólo mostramos las posiciones por las que pasa la cabeza de la máquina de Turing.

Observamos que esta máquina se detiene para todas las entradas. Aunque en los diagramas no se distingue la posición de la cabeza (se puede descubrir dónde está, pues se trata de la celda que en cada caso modifica su color), en la última fila de cada computación la cabeza sale de la cinta. No todas las máquinas de Turing se detienen para todas las entradas, como es bien conocido.

**Funciones calculadas por máquinas de Turing pequeñas**

Nuestro estudio se centra en máquinas de Turing pequeñas. Antes de definir qué son estas máquinas veamos qué se entiende por la noción de *máquina de Turing universal*. Volviendo al diagrama de la máquina 1447, si prescindimos de los pasos intermedios, vemos que a cada configuración inicial de la cinta le corresponde una configuración final, donde llega tras completar la computación. Recordemos también que las máquinas de Turing se pueden enumerar. Pues bien, la enumeración que nosotros usamos es sólo una de las formas de *codificar* las máquinas de Turing. Del mismo modo que a nuestra máquina le asignamos el número decimal 1447, podríamos asignarle un número binario o codificarla con celdas blancas y negras en la cinta de una máquina de Turing. Pues bien, una máquina de Turing $M_1$ es *universal* cuando existe una codificación que transforma cualquier máquina de Turing en una secuencia de símbolos de $M_1$ tal que para cualquier máquina de Turing $M_2$ y configuración de cinta $t$ , la ejecución de $M_1$ sobre la cinta que originalmente contiene la codificación de $M_2$ seguida de $t$ llega al mismo resultado que la ejecución de $M_2$ con entrada $t$ . Es decir, una máquina de Turing $M_1$ es universal si puede "imitar" el comportamiento de cualquier otra máquina de Turing $M_2$ . Para que $M_1$ imite el comportamiento de $M_2$ debemos pasarle la codificación de $M_2$ en los símbolos que $M_1$ es capaz de manejar. Esta codificación tiene el mismo efecto que los programas que instalamos en nuestra computadora para poder realizar diferentes tareas. Cuando la máquina universal $M_1$ lee la codificación de $M_2$ podemos decir que se configura para comportarse como $M_2$ .

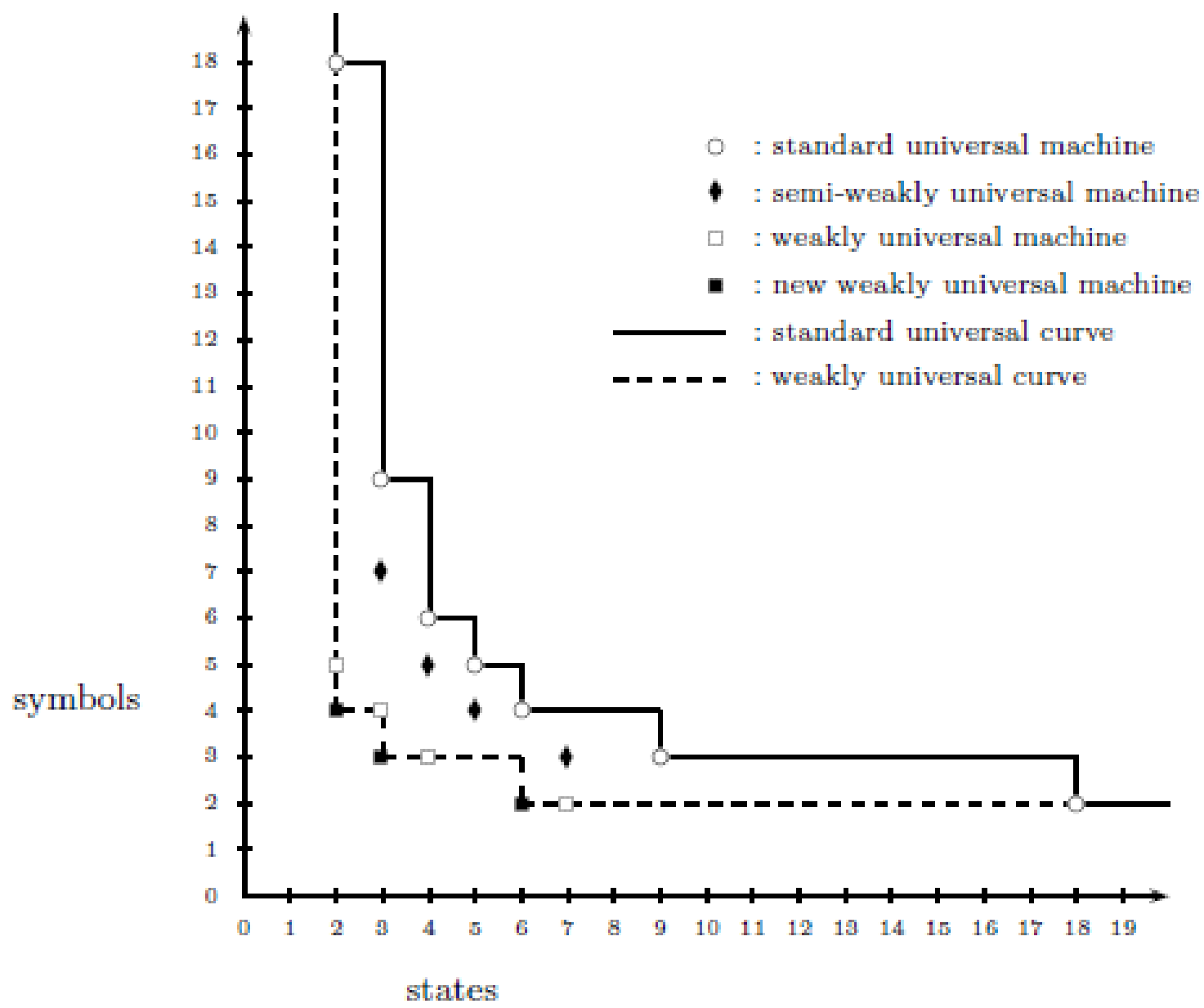

*Ilustración 3: Universalidad en máquinas de Turing pequeñas (Neary y Woods, 2007)*

El lector puede encontrar definiciones de máquinas de Turing universales en cualquier manual de teoría de la computación. A nosotros nos interesan especialmente aquellas máquinas universales que emplean menor número de estados o de símbolos. Neary y Woods (2007) ofrecen el diagrama de máquinas de Turing universales conocidas que representamos en la ilustración 3. Vemos que por debajo de cierto número de estados y símbolos no se conocen máquinas universales en ninguno de los sentidos que suele entenderse la universalidad. A este diagrama habría que añadir la máquina de Turing de 2 estados y 3 símbolos que Wolfram (2002) conjeturaba podía ser universal. Smith (2007) muestra que lo es para ciertas condiciones iniciales, resultado que, dado el tamaño de la máquina de Turing, exhibe el sistema universal más pequeño posible.

Las máquinas de Turing que estudiamos en este trabajo se encuentran en los límites conocidos de la universalidad. Aún así, hay interesantes estudios sobre complejidad que se pueden realizar con estas máquinas.

Podemos entender que cada máquina de Turing calcula una función. Dada una configuración inicial de la cinta (que puede ser la representación de un número) devuelve un valor (podemos entender que la cinta codifica un número al completarse la computación). Cuando una máquina no se detiene para cierta entrada podemos entender que la función no está definida para dicho valor.

Para elegir el formato más adecuado con que codificar en la cinta las entradas de la función, la opción más obvia, con máquinas que manejan dos símbolos, es usar codificación binaria. Sin embargo encontramos el problema que ilustramos en la siguiente imagen:

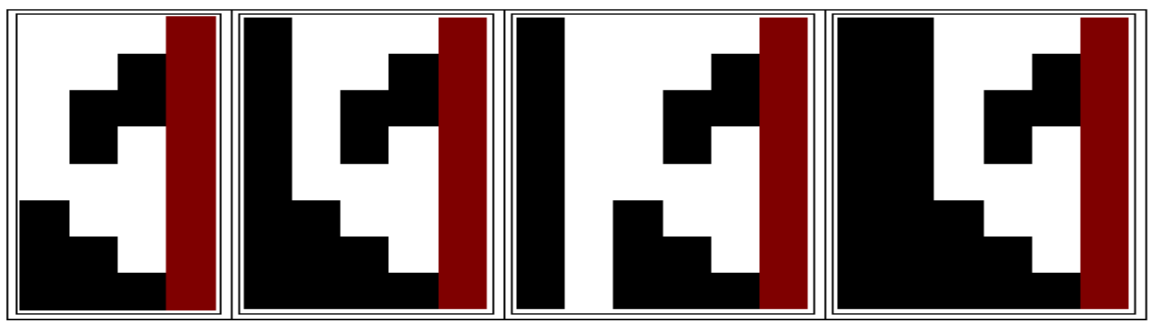

*Ilustración 4: El problema de la codificación binaria (máquina 1447)*

De izquierda a derecha, vemos las computaciones de la máquina 1447 para las entradas 0, 8, 16 y 24 (en binario). Dado que para la entrada 0 sólo se usan las tres primeras celdas de la cinta, si éstas están vacías en la entrada, la máquina no visita nunca lo que hay más allá, sea lo que sea. Es lo que ocurre en las siguientes computaciones, se han escrito diferentes entradas que sólo varían con respecto a 0 en las celdas de la cuarta en adelante, y la computación es la misma. Recuérdese que definimos la computación de una máquina de Turing como la secuencia de entradas de la tabla de transiciones que aplica, lo que es dependiente sólo de la parte de la entrada por la que pasa la cabeza. Es realidad, se puede demostrar que para cualquier máquina de Turing $M$ que pare con cierta entrada $e$ se puede reconstruir esta anomalía, dado que como máximo se visitaron $n$ celdas, por tanto toda configuración de la cinta de entrada que sea igual a $e$ en las primeras $n$ posiciones hará que $M$ realice la misma computación que para $e$. El problema más importante es que si codificamos en binario no existe ninguna máquina de Turing de dos símbolos capaz de leer cualquier entrada en binario (recorrer todas las celdas de su codificación en la cinta de entrada) y detenerse. Para verlo, supongamos que al leer la entrada 1 (primera celda negra y resto de la cinta blanca) la máquina, tras realizar las computaciones correspondientes se detiene siendo $n$ la celda más a la izquierda que se visitó. De nuevo, cualquier entrada que sea igual a 1 en los $n$ últimos dígitos hará que la máquina realice la misma computación y se detenga sin haber leído toda la entrada.

Esta situación se evita cuando la máquina dispone de al menos un símbolo de cinta más de los que se usan para codificar la entrada. Este símbolo sirve para detectar el fin de la codificación de la entrada. En tal caso, existen máquinas capaces de leer cualquier entrada completa y detenerse. Como las máquinas que manejaremos tendrán dos símbolos, lo que hacemos es usar codificación unaria con los símbolos 1 (celdas negras) y dejar el 0 (celdas blancas) como símbolo por defecto en todas las celdas no ocupadas por la entrada.

La codificación unaria o en base uno será: cero (1), uno (11), dos (111), etc. Es decir, una secuencia de $n+1$ celdas 1 (negras) para representar el número $n$. El primer 1 se sitúa en en inicio de la cinta, de modo que la máquina de Turing arranca leyéndolo.

Necesitamos también una codificación para interpretar la salida. Para no perder información del estado final de la cinta, lo más adecuado es usar, ahora sí, codificación binaria. El dígito menos significativo será el que marca el inicio de la cinta. Interpretamos toda la cinta (claro está, hasta el 1 de más a la izquierda, que será el dígito más significativo), independientemente de que la máquina haya visitado o no todas las celdas.

Con esta codificación sólo debemos tener en cuenta que un gran número de máquinas, entre ellas muchas triviales que terminan dejando la cinta tal como estaba, van a calcular funciones aparentemente exponenciales, pero ello se debe sólo a la distinta convención usada para la entrada y la salida. Observemos:

| Valor de entrada | Cinta | Valor de salida |
|---|---|---|
| 0 | 1 | 1 |
| 1 | 11 | 3 |
| 2 | 111 | 7 |
| 3 | 1111 | 15 |

Por tanto una máquina de Turing que termine dejando la cinta tal como estaba inicialmente computará la función $2^{n+1}-1$ (en ocasiones en tiempo constante) pero no debemos engañarnos, pues se debe a las convenciones de entrada/salida elegidas.

Nuestro trabajo consiste en explorar distintos espacios de máquinas de Turing con dos símbolos y distintos números de estados y estudiar qué funciones se pueden calcular en cada uno de tales espacios. Estamos especialmente interesados en cuál es el coste en tiempo (número de pasos de computación) y espacio (número de celdas de la cinta usadas durante la computación) que requiere calcular cada función según el número de estados disponibles. Nuestra hipótesis es que ciertas funciones podrán ser calculadas con un menor coste computacional al aumentar el número de estados, aunque el tiempo y espacio medio requeridos para calcular una función aumentará con el número de estados de las máquinas.

**Metodología**

Vamos a centrarnos en máquinas de Turing con dos símbolos y distintos números de estados. Recordemos que el número de entradas de la tabla de transiciones de una máquina con dos símbolos y $n$ estados es $2n$ . Cada transición tiene la forma: (nuevo símbolo, nuevo estado, movimiento). Dado que hay 2 símbolos, $n$ estados y 2 movimientos (izquierda o derecha), hay $2*n*2=4n$ posibles transiciones en cada una de las $2n$ entradas. Esto nos da un total de $4n^{2n}$ máquinas de Turing diferentes con 2 símbolos y $n$ estados. La siguiente tabla muestra cómo crece este número:

| **Estados** $n$ | **Máquinas** $4n^{2n}$ |
|---|---|
| 1 | 16 |
| 2 | 4096 |
| 3 | 2985984 |
| 4 | 4294967296 |

En adelante, usaremos $(s,k)$ para referirnos al conjunto de máquinas con $s$ estados y $k$ símbolos. Para $(2,2)$ y $(3,2)$ hemos podido ejecutar todas las máquinas de Turing con entradas desde 0 hasta 20 (en notación unaria, recordemos). Para ello hemos dispuesto de los recursos de supercomputación del Centro de Informática Científica de Andalucía (CICA), donde ejecutamos un simulador programado en C. Todas las máquinas que no terminaron antes de 1000 pasos se cortaron. Posteriormente se aplicaron distintos procedimientos para completar las computaciones que terminaron con más de 1000 pasos. Todo el análisis de datos se realizó utilizando el software *Mathematica* de Wolfram Research.

La exploración de $(2,2)$ requirió media hora y se utilizó un solo procesador del CICA.

Para $(3,2)$ necesitamos 25 procesadores trabajando en simultáneo durante unas 3 horas (alrededor de 75 horas en total). Con la misma velocidad de cálculo, la exploración completa de $(4,2)$ requiere poco más de 12 años. El mayor problema es que mientras que toda la salida de la exploración de $(3,2)$ ocupó alrededor de 2 Gigabytes, la de $(4,2)$ requeriría unos 3 Terabytes, por lo que hoy en día nos resultaría imposible extraer información de tal cantidad de datos. Por tanto, para $(4,2)$ lo que hacemos son muestreos de varios millones de máquinas para poder tener resultados que comparar con los de $(3,2)$.

Merece la pena comentar una de las técnicas que hemos utilizado en el procesamiento de datos de $(3,2)$ para evitar duplicar ciertas computaciones. En este espacio hay pares de máquinas de Turing que podemos llamar *gemelas*, pues aunque sus tablas de transiciones no son idénticas, existe cierta simetría entre ellas. Estas máquinas realizan computaciones equivalentes y, para cada entrada, o ambas paran y llegan a la misma salida, o no lo hace ninguna de las dos. Por ejemplo, veamos las tablas de transiciones de las dos máquinas siguientes:

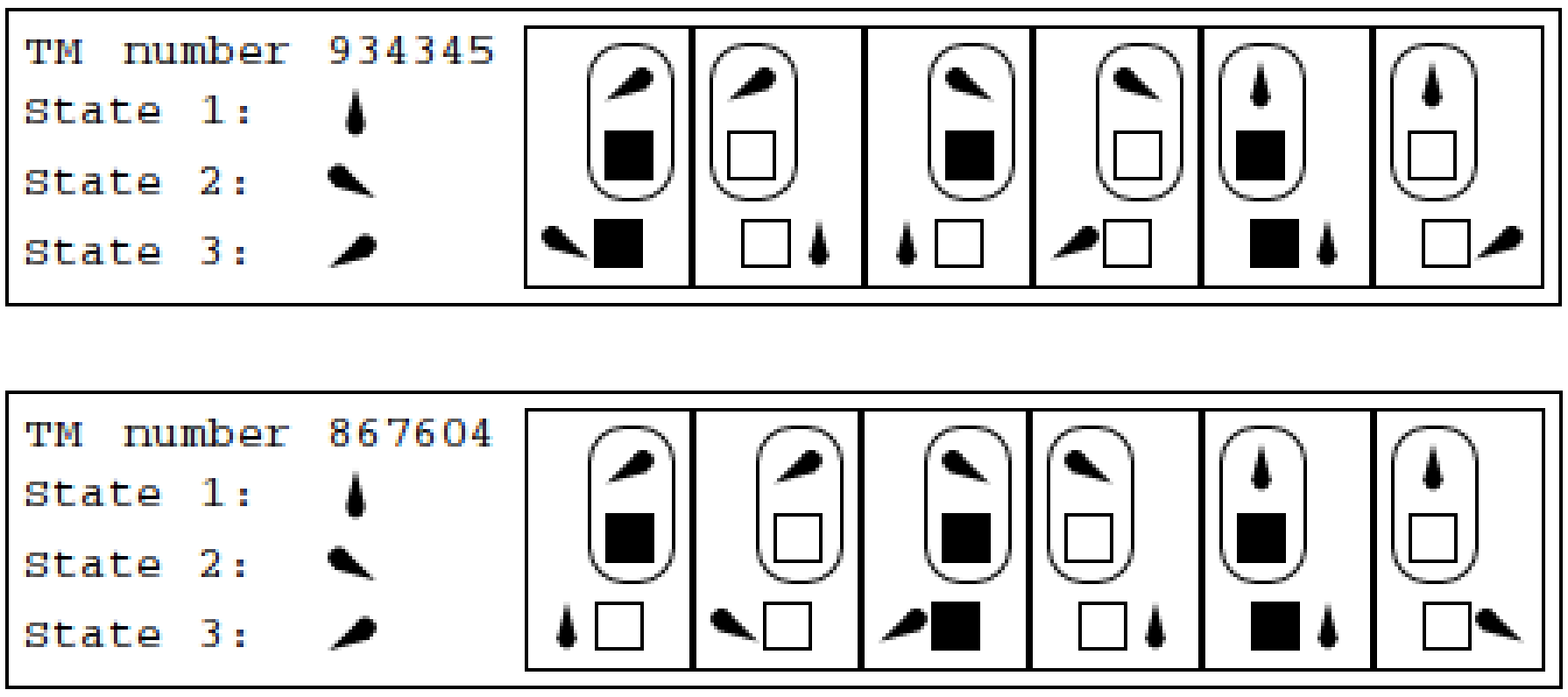

*Ilustración 5: Máquinas gemelas*

Cada una de estas reglas se puede obtener a partir de la otra permutando el comportamiento de los estados 2 y 3 (los no iniciales). Por ejemplo, en 934345 hay una transición desde el estado 1 al 3 y un movimiento a la derecha cuando se lee una celda blanca. En 867604 la transición simétrica es al estado 2. También, en 934345 cuando la máquina está en el estado 2 y lee un blanco se mueve a la izquierda y cambia al estado 3. Ahora en 867604 se hace lo simétrico al leer un blanco en el estado 3: también se mueve a la izquierda pero ahora cambia al estado 2. Esto es lo que pasa en máquinas gemelas: el comportamiento de los estados 2 y 3 está permutado, por lo que hacen computaciones equivalentes (sólo cambia el número de los estados) para cualquier entrada.

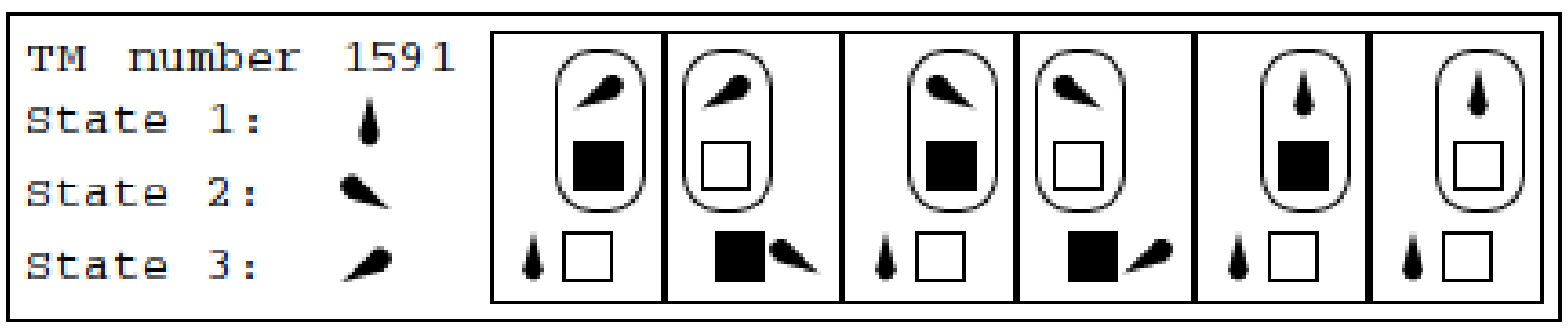

*Ilustración 6: Una máquina gemela de sí misma*

No para todas las máquinas de Turing en $(3,2)$ existe una máquina gemela diferente. Veamos por ejemplo la máquina representada en la ilustración 6. La misma simetría que explicamos para las máquinas anteriores se da ahora entre esta máquina y ella misma. Una característica de las máquinas gemelas de sí mismas es que los estados 2 y 3 nunca se alcanzan, se puede observar que no hay transiciones desde el estado 1 ni al 2 ni al 3. En otro caso, si hubiera una transición desde el estado 1 a alguno de los otros, al intercambiar los estados 2 y 3, la máquina gemela no podría ser ella misma, al obtenerse una transición diferente. Por tanto estas máquinas usan en realidad un solo estado. Para alguno de los análisis que se han realizado con los datos de $(3,2)$ se han tenido en cuenta estas simetrías y por cada par de máquinas gemelas diferentes sólo una de ellas se ha considerado en los cálculos.

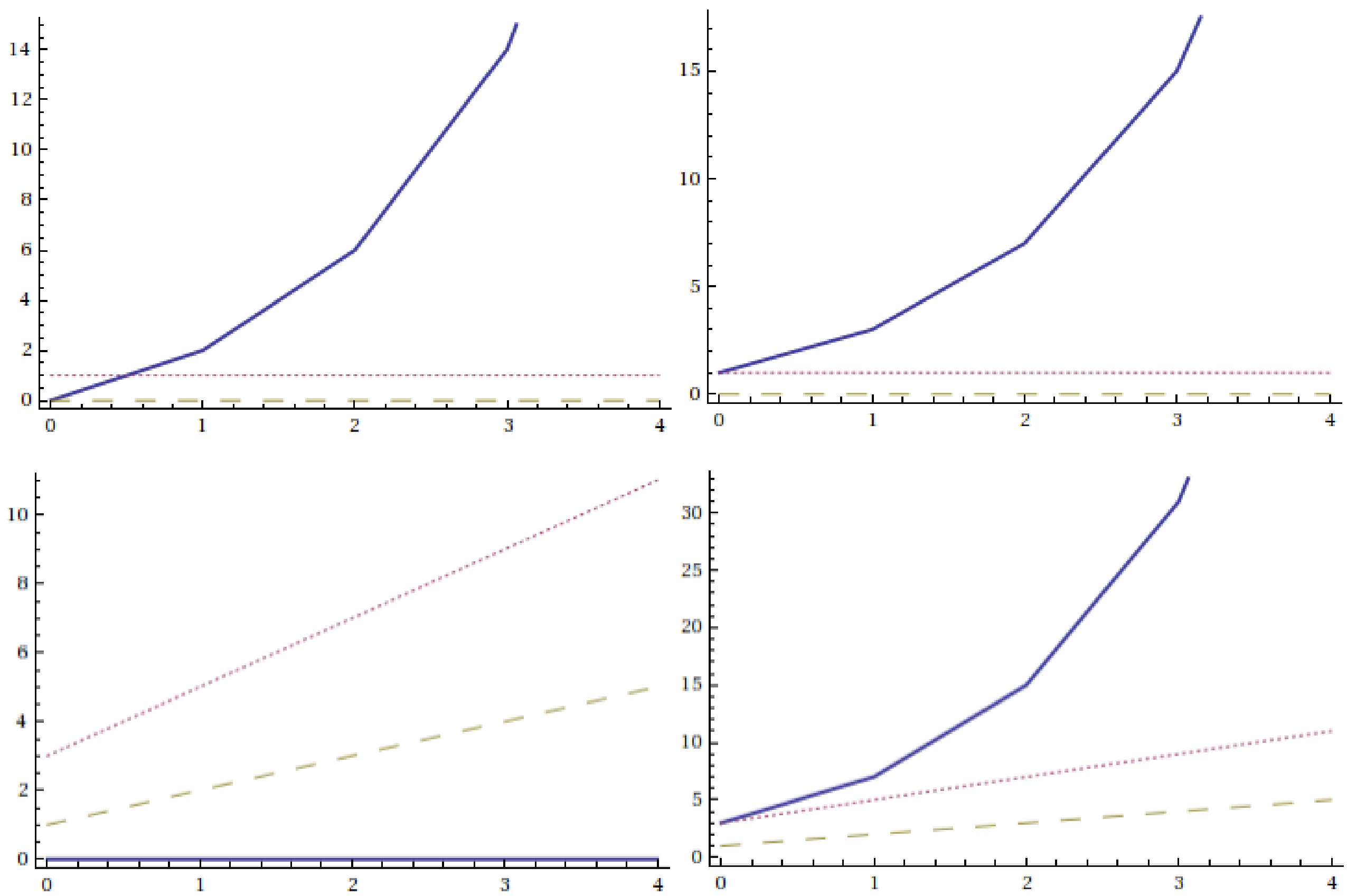

*Ilustración 7: Funciones calculadas en (1, 2)*

**Exploración de (1,2)**

El espacio más simple, pero no por ello falto de importancia, es el de máquinas con un solo estado y dos símbolos, lo que representamos como $(1,2)$. Hay un total de 16 máquinas. De ellas hay 6 que no paran para ninguna de las entradas. El la ilustración 7 vemos las funciones que calculan las otras 10. En realidad, los gráficos que mostramos representan más bien *algoritmos* (funciones a las que se les asocian costes de tiempo y espacio). En espacios mayores a $(1,2)$ tendremos que hacer la diferencia entre algoritmo y función, ya que habrá funciones que son calculadas con diferentes algoritmos, pero no es el caso ahora, por lo que usaremos indistintamente ambos términos.

La ilustración 7 muestra las funciones calculadas por máquinas de $(1,2)$ que siempre se detienen. Hay cuatro máquinas que calculan cada una de las dos funciones superiores y una que calcula cada una de las dos inferiores. En cada caso la línea gruesa representa la salida de la función (en binario, como explicamos), la línea punteada el coste en tiempo y la rayada el coste en espacio. Sólo se han representado los cinco primero valores de cada función.

Las dos funciones superiores se puede observar que terminan siempre con tiempo 1 (esto es, un paso de computación) y espacio 0 (la posición 0 de la cinta es la inicial). Corresponden a máquinas de Turing que en el estado inicial, al leer un 1, se desplazan a la derecha, y por tanto salen de la cinta y terminan la computación. Si en esta única transición escriben un 1 dejan la cinta como estaba y computan la identidad (superior derecha). Si escriben un 0, restan una unidad a la identidad (superior izquierda). En cualquier espacio $(n,2)$ mediante simples cálculos combinatorios sabemos que:

- El 25% de las máquinas calculan el algoritmo superior izquierdo, esto es, al leer negro en el estado inicial escriben blanco y mueven a la derecha (sea cual sea el estado al que pasen, salen de la cinta).
- El 25% de las máquinas calculan el algoritmo superior izquierdo, esto es, hacen lo mismo que las anteriores pero dejan el negro que estaba escrito.

Los dos gráficos de abajo tienen un coste lineal en tiempo y espacio. En ambos casos la secuencia de tiempos es {3, 5, 7, 9, 11, … } y la de espacios {1, 2, 3, 4, 5, … }. Si representáramos las computaciones, son máquinas que recorren toda la entrada completa y al llegar al final (primera celda blanca) vuelven al inicio y terminan. En el caso de la izquierda, la máquina recorre la entrada borrando los negros, y al encontrar el primer blanco, lo deja sin modificar y se vuelve, dejando sin alterar las celdas blancas que encuentra. Por eso la salida siempre es 0 (la cinta queda totalmente blanca al terminar). La máquina de la derecha recorre la entrada también borrando los negros, pero al llegar al primer blanco vuelve y tanto ese blanco como los que va encontrando a la vuelta los va cambiando a negro, por lo que la salida es como la entrada con un negro más al final. Podemos resumir las reglas que siguen ambas máquinas:

| | **Máquina de la izq.** | **Máquina de la der.** |
|---|---|---|
| **Encuentra negro** | Escribe blanco, mueve izq. | Escribe blanco, mueve izq. |
| **Encuentra blanco** | Escribe blanco, mueve der. | Escribe negro, mueve der. |

Pese a que las funciones que hemos encontrado en $(1,2)$ son muy simples, vemos que se calculan con el menor coste posible en tiempo y espacio. Por tanto, el coste medio de calcularlas en $(2,2)$ y $(3,2)$ aumentará. Son, pues, casos donde sabemos que sólo podemos encontrar aumento del coste promedio, y ninguna máquina que mejore el coste de $(1,2)$.

**Exploración de (2,2)**

En $(2,2)$ hemos encontrado 74 funciones que son calculadas por 253 algoritmos. Una indicación de complejidad es el número de valores necesarios para determinar una función. En este caso bastan los tres primeros. Para la primera entrada hay 11 salidas diferentes, y para las dos primeras 55 combinaciones distintas. La primera salida de las funciones, correspondiente a la entrada 0, es determinante, ya que si no la consideramos, en $(2,2)$ sólo hay 45 combinaciones del resto de valores. Esto se debe a que hay casos de funciones que sólo difieren en la primera salida.

El siguiente gráfico muestra la distribución de probabilidad de detención de todas las máquinas de Turing de $(2,2)$ considerando todas las entradas:

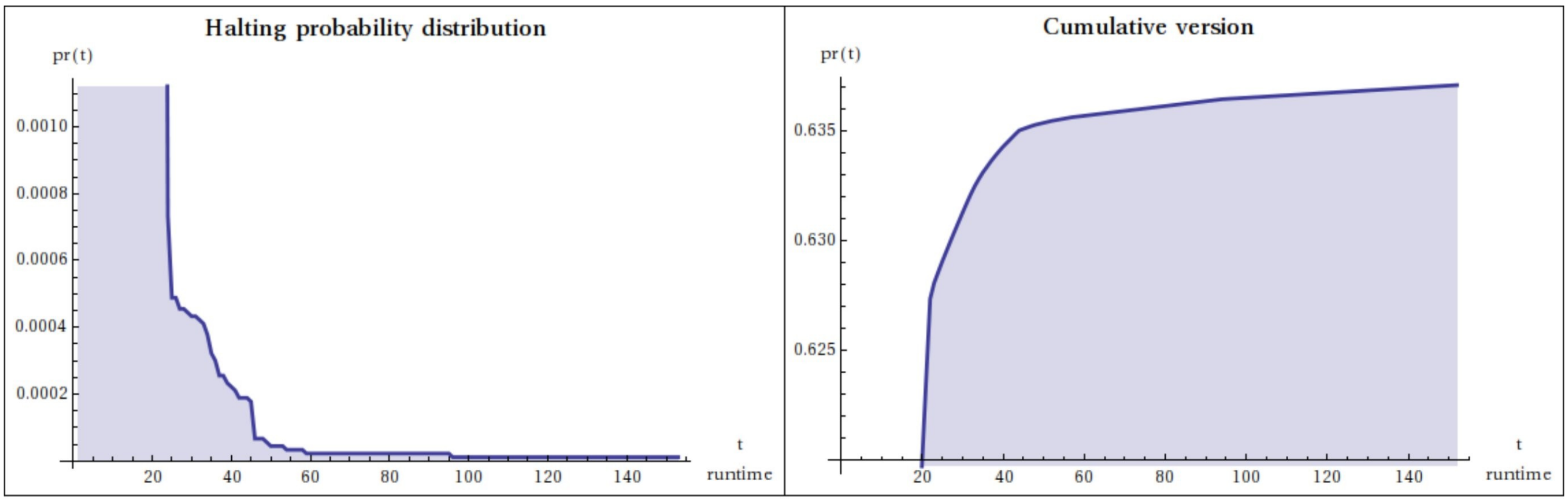

*Ilustración 8: Probabilidad de detención en (2,2)*

En horizontal tenemos el número de pasos y en vertical la probabilidad de que una máquina de Turing aleatoria de $(2,2)$ con una entrada aleatoria entre 0 y 20 se detenga en dicho número de pasos. Si vemos el gráfico de la derecha observamos que algo más del 63% de las ejecuciones terminan en menos de 50 pasos. A partir de ahí el crecimiento es mínimo. Considerando que hay cierta cantidad de ejecuciones que nunca paran, confirma la hipótesis de Calude (2005) que afirma que la mayoría de máquinas de Turing paran rápidamente o nunca lo hacen.

Hay un total de 49 secuencias de tiempos de ejecución, 35 si excluimos funciones no totales. El mayor tiempo de ejecución ocurre en las máquinas número 378 y 1351 que corren 8388605 para la última entrada. Para la máquina 378 esta es la secuencia de {entrada, salida, tiempo, espacio}:

{{0, 1, 5, 1}, {1, 3, 13, 2}, {2, 7, 29, 3}, {3, 15, 61, 4}, {4, 31, 125, 5}, {5, 63, 253, 6}, {6, 127, 509, 7}, {7, 255, 1021, 8}, {8, 511, 2045, 9}, {9, 1023, 4093, 10}, {10, 2047, 8189, 11}, {11, 4095, 16381, 12}, {12, 8191, 32765, 13}, {13, 16383, 65533, 14}, {14, 32767, 131069, 15}, {15, 65535, 262141, 16}, {16, 131071, 524285, 17}, {17, 262143, 1048573, 18}, {18, 524287, 2097149, 19}, {19, 1048575, 4194301, 20}, {20, 2097151, 8388605, 21}}.

En la ilustración 9 vemos la información correspondiente a la función a la que pertenece esta máquina. Cada una de las filas de la matriz superior muestra el estado de la cinta de la máquina de Turing al terminar una computación. Así, la fila superior muestra la cinta al completar la computación para la entrada 0 y la inferior para la entrada 20. Como vemos, es la función identidad, ya que en cada caso devuelve la cinta tal como estaba originalmente.

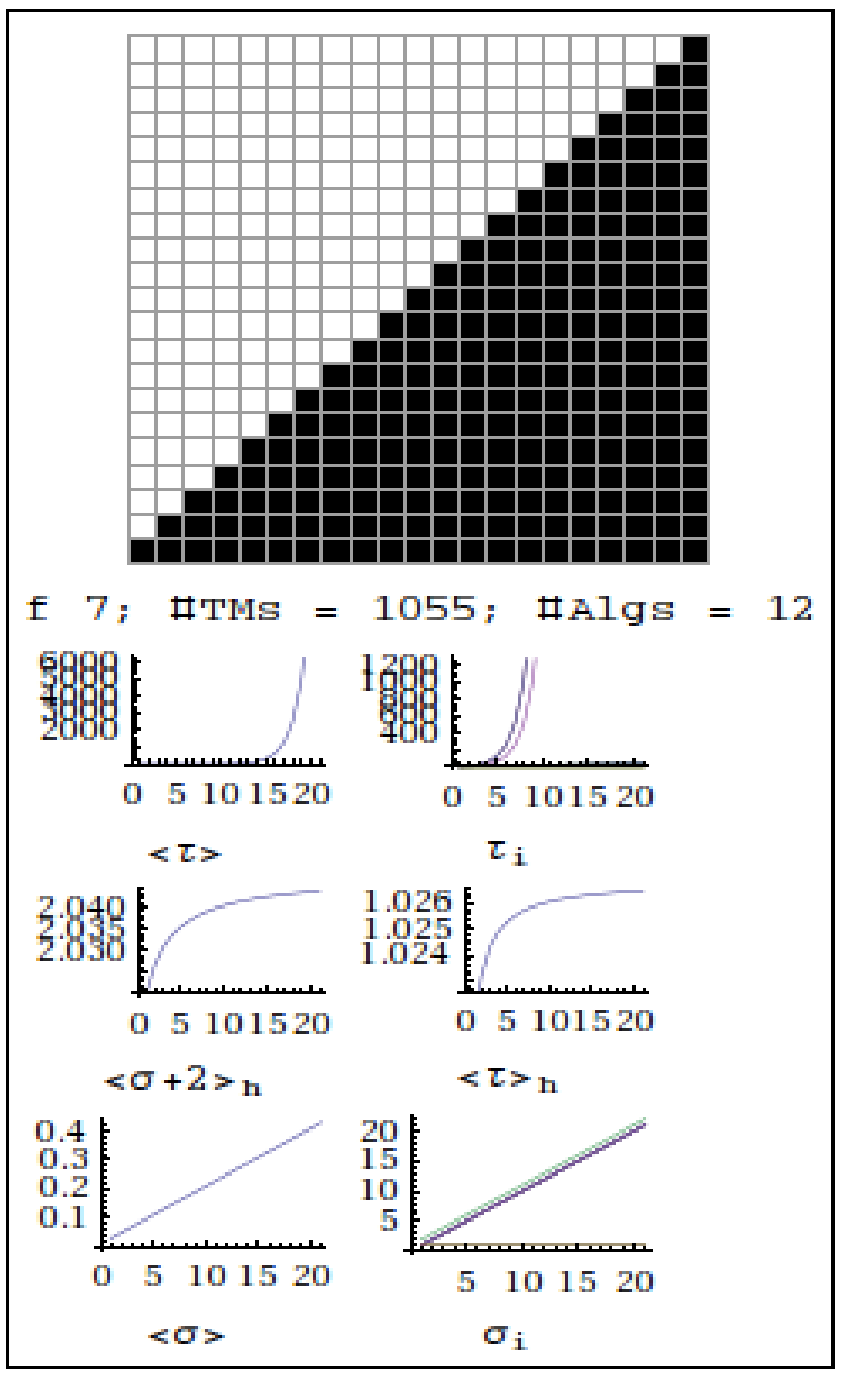

*Ilustración 9: Identidad en (2,2)*

La información inferior nos indica que esta función está etiquetada como la 7 y que es calculada por 1055 máquinas de $(2,2)$ agrupadas en 12 algoritmos. Luego vemos representados el crecimiento en tiempo $\tau_1$ y espacio $\sigma_i$ para cada algoritmo. Las medias aritméticas aparecen como $\langle\tau\rangle$ y $\langle\sigma\rangle$, respectivamente. Las medias armónicas aparecen representadas como $\langle\tau\rangle_h$ y $\langle\sigma+2\rangle_h$ (por razones técnicas, como la media armónica sólo está definida para valores distintos a cero, usamos $\sigma+2$ y no $\sigma$ ).

La media o promedio armónico tiene una interpretación muy útil cuando se le relaciona con máquinas de Turing. Dos máquinas de Turing que computan la misma función pueden verse como máquinas computando la misma cantidad de información aunque les pueda llevar distintos tiempos de ejecución. El tiempo de ejecución de una máquina en particular para una entrada en particular puede entonces ser interpretado como tiempo de ejecución/información. Si consideramos la siguiente situación: Sean las máquinas de Turing computando una misma función $M_1,\ldots,M_n$ con tiempos de ejecución $t_1,\ldots,t_n$ . Si dejamos correr cada máquina por una sola unidad de tiempo, entonces la cantidad de información computada al final es $1/t_1+\ldots+1/t_n$ . Y el promedio correspondiente a la cantidad típica de información que le lleva a una máquina de Turing calcular por unidad de tiempo es precisamente la media armónica.

Vemos que el tiempo medio crece exponencialmente mientras el espacio lo hace linealmente. Este fenómeno es bastante frecuente. Recordemos que esta función ya apareció en $(1,2)$ y comentamos que en cualquier espacio, como ocurría entonces, un 25% de las máquinas la calculan en un solo paso y sin usar más celdas de la cinta que la inicial. Así que de las 1055 máquinas de $(2,2)$ hay 1024 de este tipo. El resto son máquinas que requieren más recursos.

Veamos los dos gráficos de la ilustración 10:

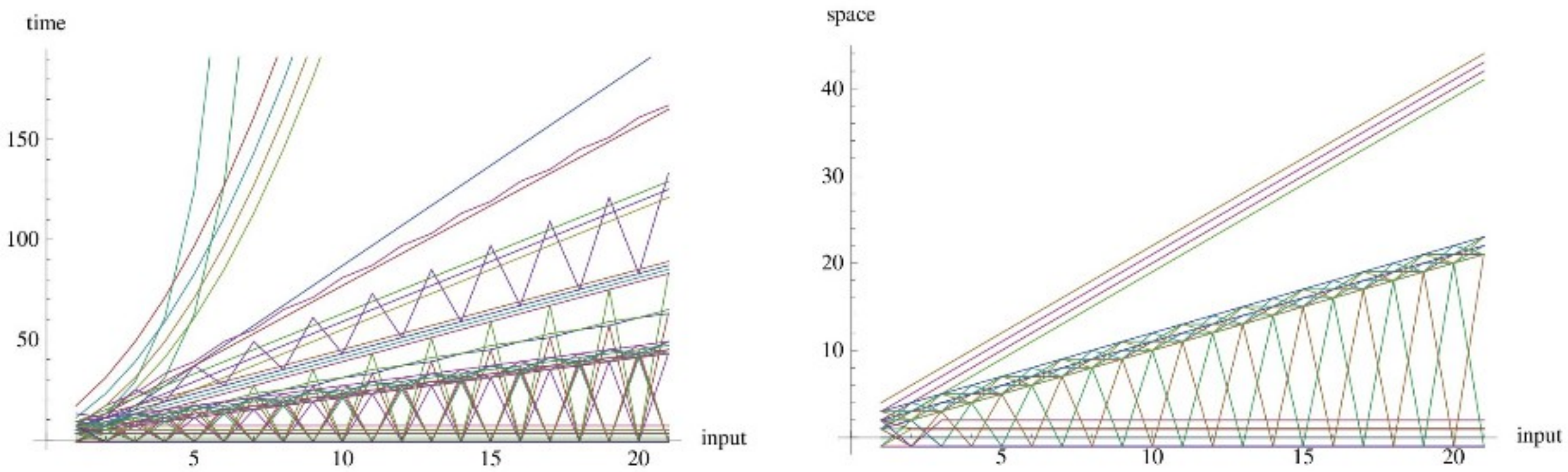

*Ilustración 10: Secuencias de tiempo y espacio en (2,2)*

El de la izquierda muestra todas las secuencias de tiempo y el de la derecha las de espacio. Los casos de divergencia (ejecuciones que no paran) se representan por -1, lo que explica los valores que bajan del eje horizontal. Encontramos algunos tiempos exponenciales pero la mayoría de ellos, al igual que el espacio, son lineales. Se observa un fenómeno interesante, casos de divergencia alternante que también hemos encontrado en otros espacios. Se trata de máquinas capaces de detenerse en entradas pares, por ejemplo, pero no en las impares.

Terminaremos este análisis con algunos comentarios sobre el tipo de computaciones que encontramos en $(2,2)$. La mayoría de máquinas hacen computaciones muy simples. El patrón que siguen, por lo general, es (aparte del 50% de máquinas que terminan en un solo paso) recorrer la cinta de entrada una sola vez. Es el caso de la máquina 2240 con la siguiente secuencia de tiempos: {5, 5, 9, 9, 13, 13, 17, 17, 21, 21, ..}. Veamos las computaciones para las entradas desde 0 a 5 (ilustración 11):

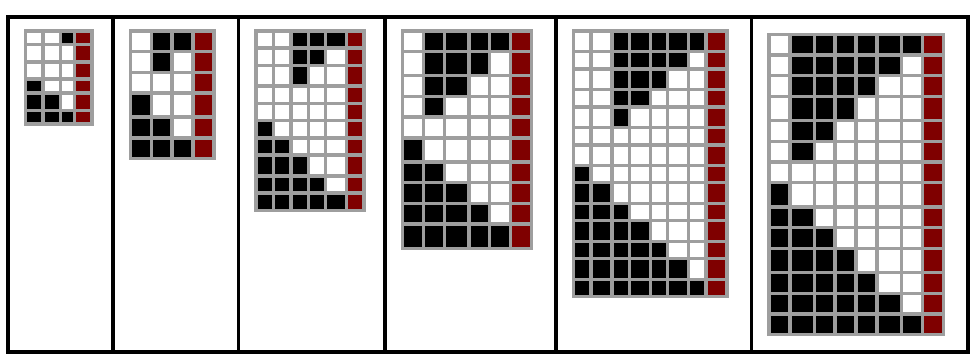

*Ilustración 11: Computaciones de la máquina 2240*

El recorrido a través de la cinta puede ser más complicado. Es el caso de la máquina 2205 con la secuencia de tiempos: {3, 7, 17, 27, 37, 47, 57, 67, 77, …}. Son mayores tiempos de ejecución pero sólo se recorre la cinta una vez, como podemos ver en las computaciones (ilustración 12). El caso de la máquina 1351 es una de las pocas que escapan de este comportamiento tan simple. Como vimos, tiene el mayor tiempo de ejecución. Viendo las computaciones se puede observar que se trata de una forma muy compleja de calcular la identidad (ilustración 13).

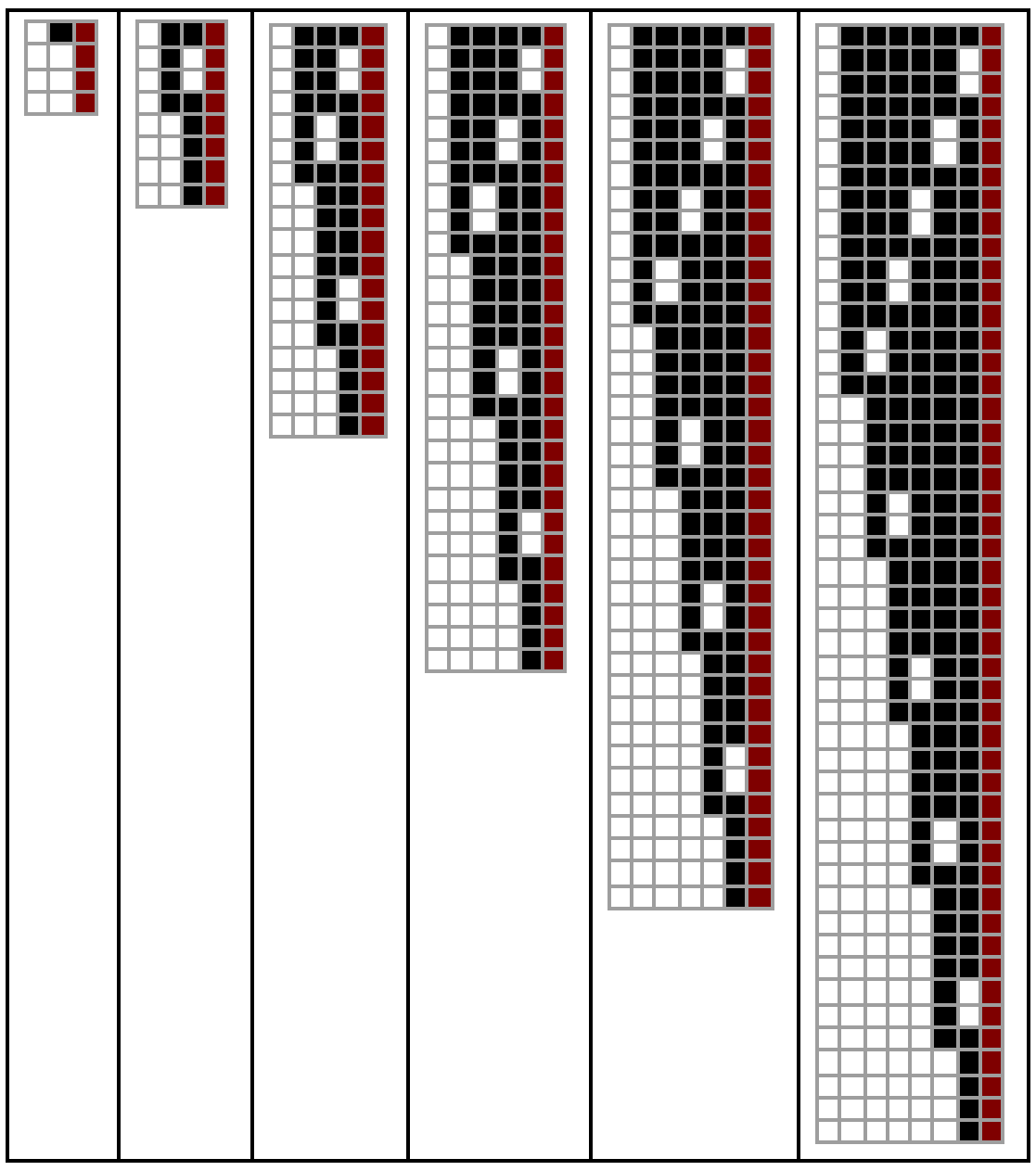

*Ilustración 12: Computaciones de la máquina 2205*

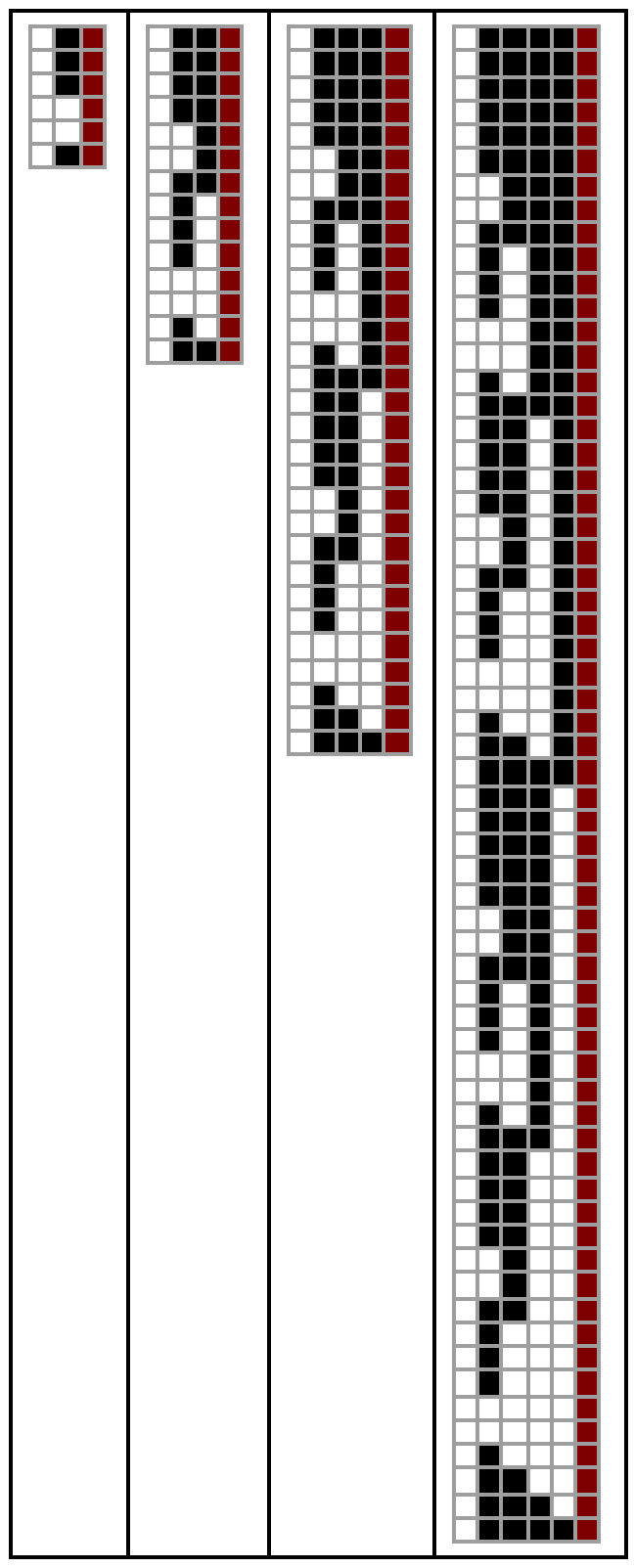

*Ilustración 13: Computaciones de la máquina 1351*

**Exploración de (3,2)**

En $(3,2)$ hay 2985984 máquinas de Turing que calculan 3886 funciones diferentes siguiendo 12824 algoritmos distintos. Como estas máquinas son más complejas que las de $(2,2)$, necesitamos más salidas para caracterizar una función. Ahora pasamos de las 3 a las 8 primeras. Veamos:

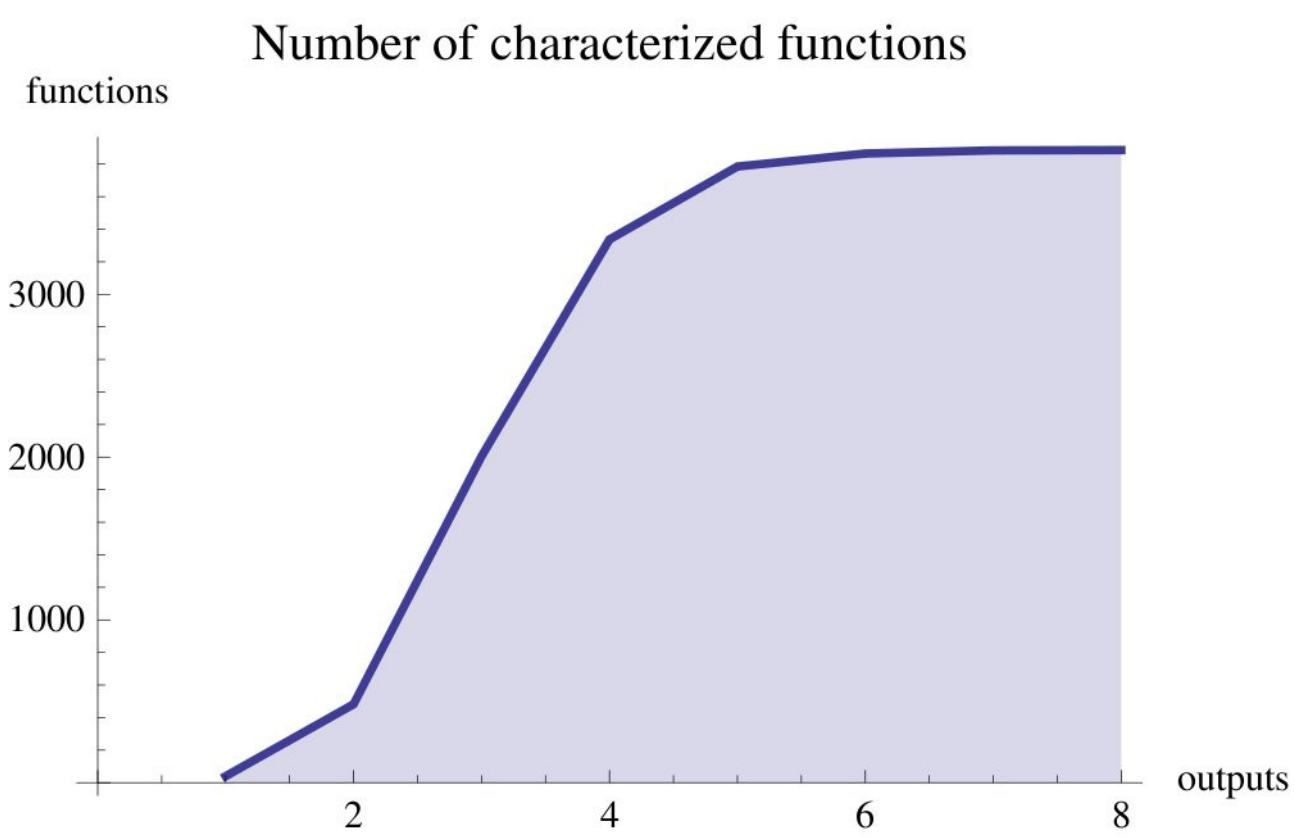

*Ilustración 14: Salidas que caracterizan una función en (3,2)*

La distribución de probabilidad del tiempo de ejecución no difiere mucho de la que encontramos anteriormente. Casi la totalidad de las computaciones que se detienen lo hacen también ahora en menos de 50 pasos (ilustración 15):

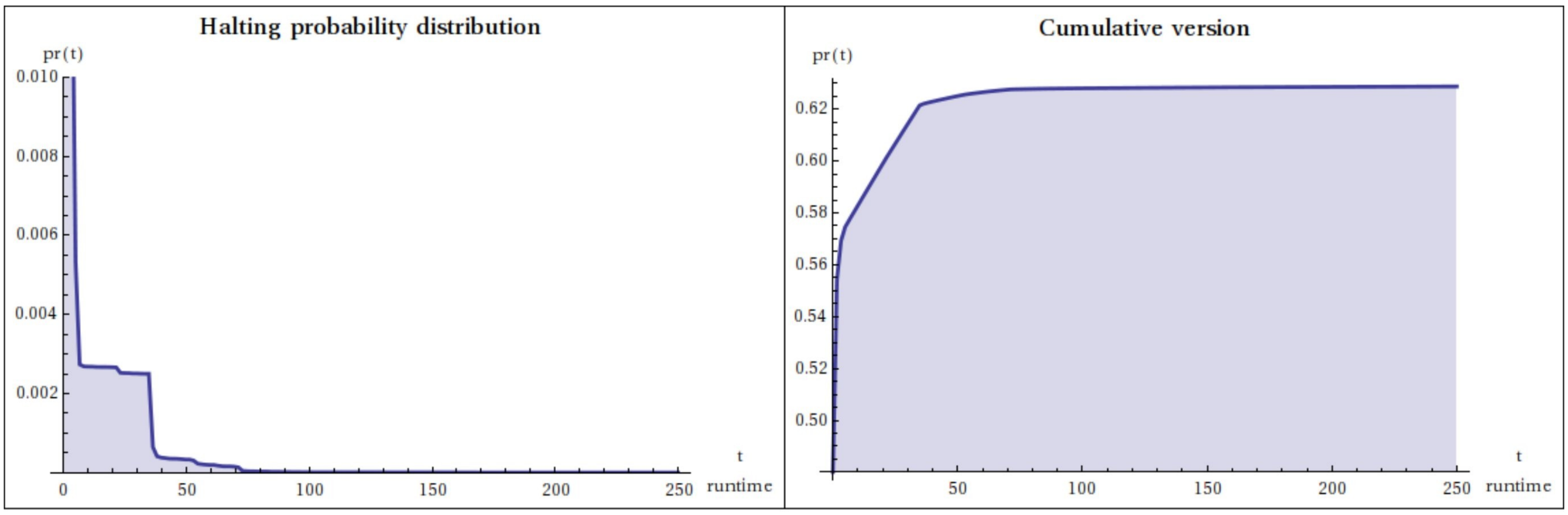

*Ilustración 15: Probabilidad de detención en (3,2)*

No podemos mostrar todas las secuencias de tiempos de ejecución y espacio porque son excesivas. Pero lo hacemos para una muestra aleatoria de 50 secuencias en cada caso (ilustración 16; tiempo a la izquierda y espacio a la derecha). Una observación interesante es que las secuencias parecen estar agrupadas:

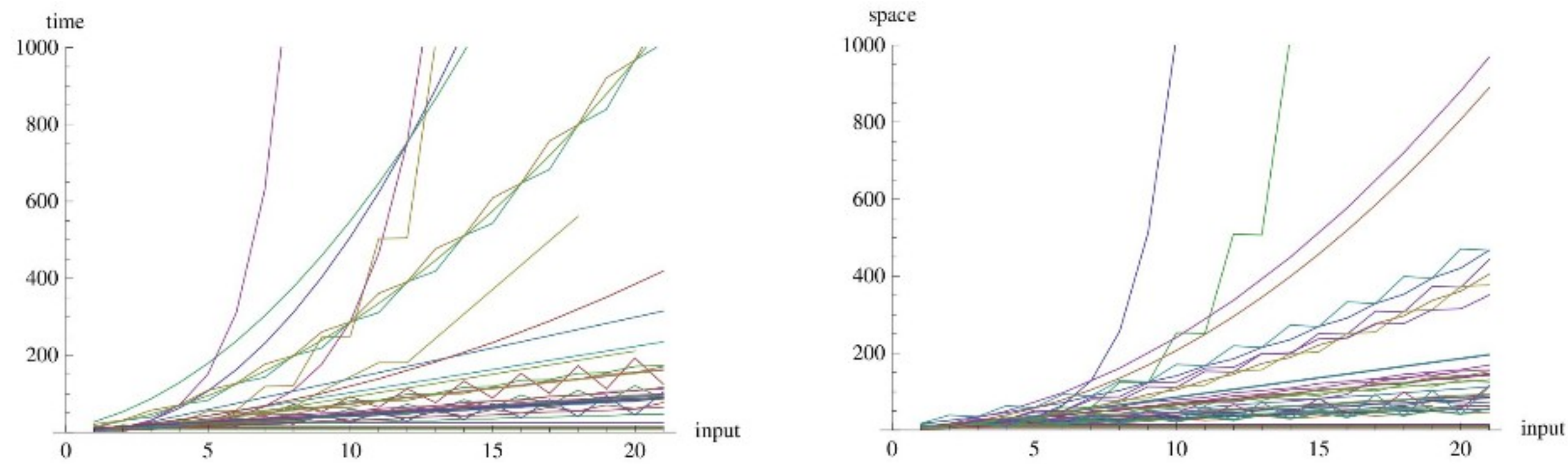

*Ilustración 16: Muestra de las secuencias de tiempos y espacios en (3,2)*

Recordemos que en $(2,2)$ la mayoría de máquinas que se detienen lo hacen en tiempo lineal. Ahora vemos más casos de comportamiento exponencial, y no sólo en el tiempo sino también en el espacio.

El máximo tiempo en $(3,2)$ son 894481409 pasos que las máquinas número 599063 y 666364 (un par de máquinas gemelas) requieren para detenerse en la entrada 20. Los valores de esta función son doblemente exponenciales. Todos ellos son iguales a una potencia de 2 a la que se restan 2 unidades. Veamos las primeras salidas: {14, 254, 16382, 8388606, 137438953470, ... } . Si sumamos 2 a cada valor y tomamos el logaritmo en base 2, tenemos: {4, 8, 14, 23, 37, 58, 89, 136, 206, 311, 469, 706, 1061, 1594, 2393, 3592, 5390, 8087, 12133, 18202, 27305} . Lo vemos

representado gráficamente:

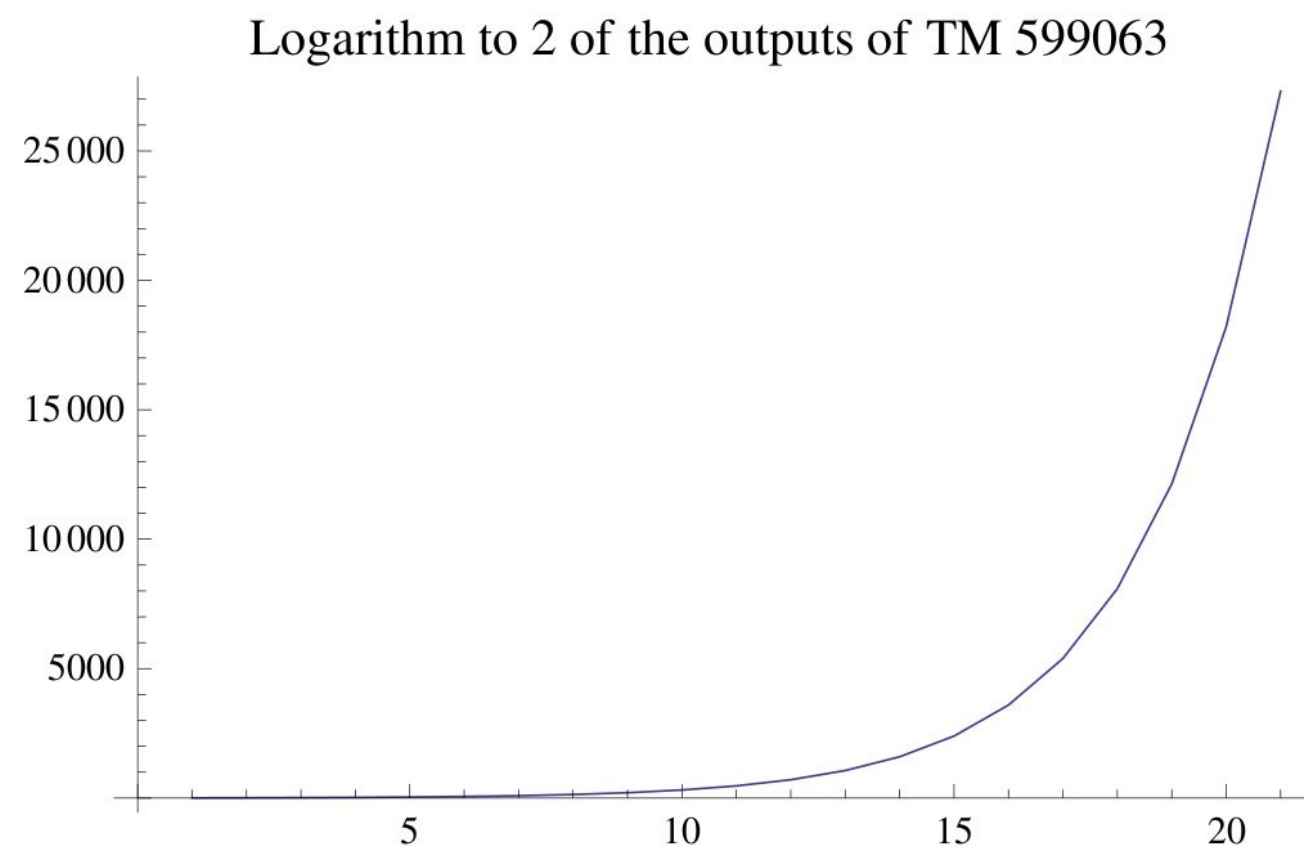

*Ilustración 17: Salida de la máquina 599063*

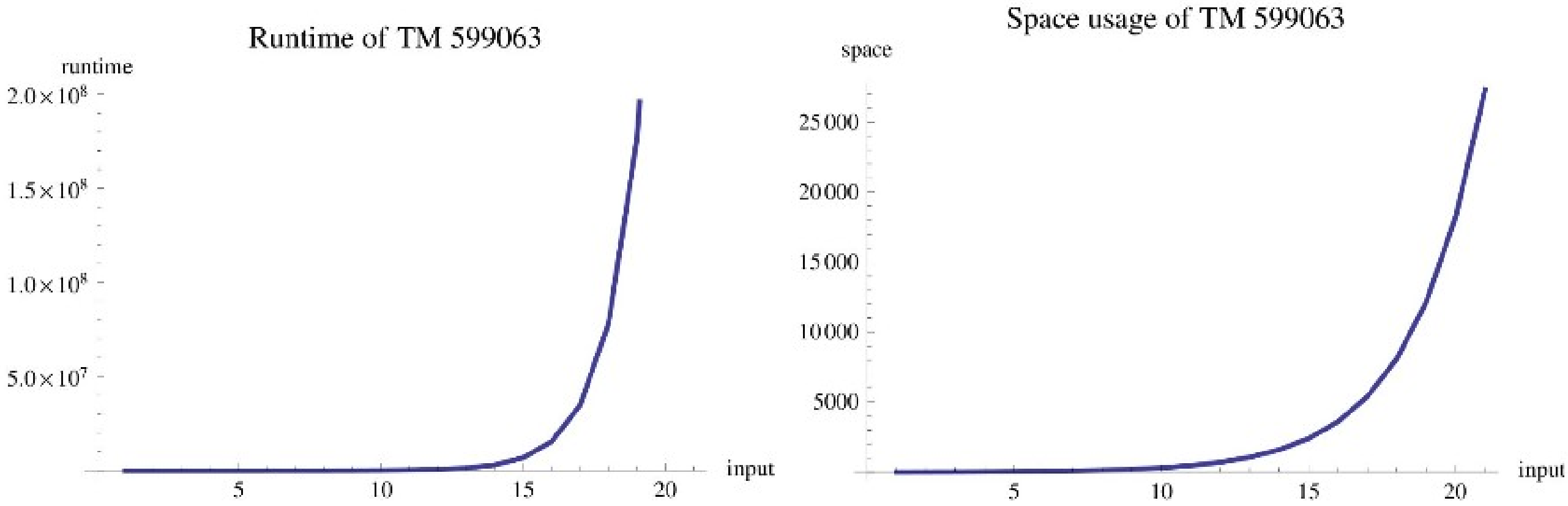

*Ilustración 18: Crecimiento en tiempo y espacio de la máquina 599063*

La ilustración 18 muestra el crecimiento en tiempo y espacio.

Recordemos que nuestra convención de salida tenía el problema de producir funciones aparentemente exponenciales cuando en realidad se estaban realizando computaciones muy simples, incluso la identidad. Ahora, al encontrar comportamiento exponencial una vez tomado el logaritmo en base 2, sí que tenemos una máquina que realiza una computación típicamente exponencial.

Podemos ver en la ilustración 19 algunas computaciones. El patrón que observamos en la segunda de ellas se repite un número exponencial de veces en las sucesivas entradas.

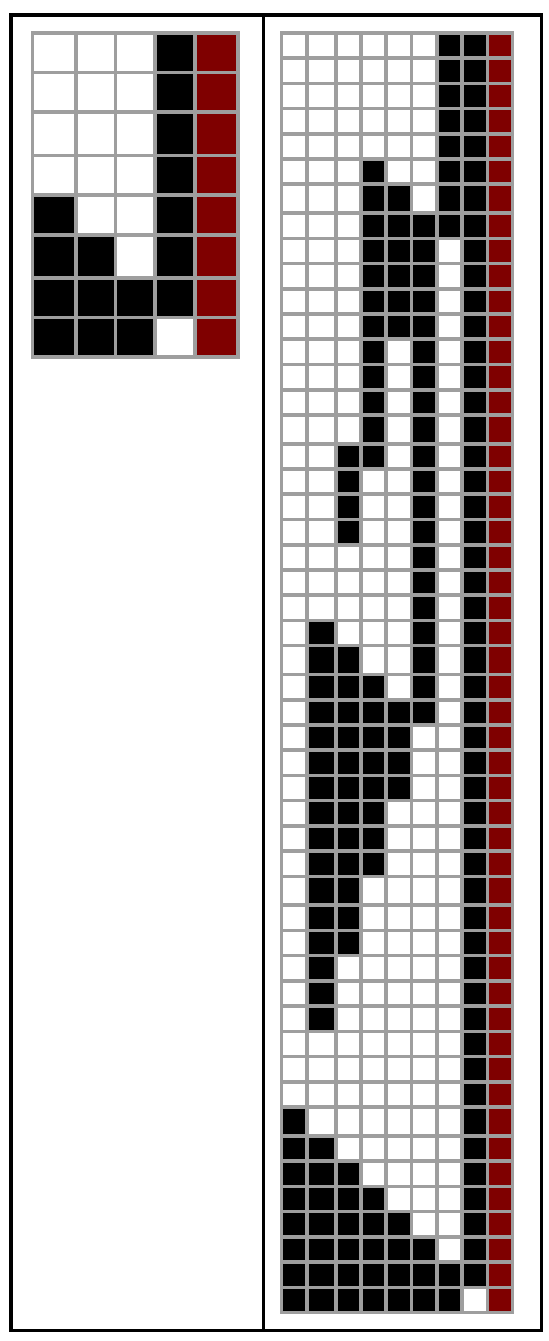

*Ilustración 19: Computaciones de la máquina 599063*

**Conclusiones**

Al iniciar el proyecto al que pertenecen los resultados aquí mostrados, partíamos de la hipótesis de que encontraríamos un número significativo de casos en los que una función sería computada con un menor tiempo medio al aumentar el número de estados. Es decir, casos donde las máquinas con $n+1$ estados que calculan la función $f$ lo hacen, en promedio, más rápido que las que calculan $f$ con $n$ estados. Lo que hemos encontrado, sin embargo, es que por lo general aumenta el tiempo medio de computación para calcular $f$ al aumentar el número de estados. Por tanto, las afirmaciones que podemos hacer en el estado actual de nuestro proyecto son:

1. Todo lo que se puede hacer con $n$ estados se puede también con $n+1$ estados. Esto es un hecho, dado que en $(n+1,2)$ están trivialmente incluidas todas las máquinas de $(n,2)$, pues basta con no definir ninguna transición hasta el último estado desde los anteriores. Por tanto, todos los algoritmos que encontramos en $(n,2)$ aparecen también en $(n+1,2)$.
2. En ciertos casos, aparecen en $(n+1,2)$ algoritmos más eficientes (menor tiempo y/o espacio) que los existentes en $(n,2)$ para calcular una misma función $f$.
3. Por lo general, dada una función $f$, los algoritmos que la calculan en $(n+1,2)$ tienen un peor rendimiento medio que los que la calculan en $(n,2)$.

Las observaciones 2 y 3 no son contradictorias. La 2 nos dice que en $(n+1,2)$ puede haber *algún* algoritmo mejor a todos los presentes en $(n,2)$ y la 3 afirma que por lo general el coste *medio de todos* los algoritmos de $(n+1,2)$ es mayor que es de los algoritmos de $(n,2)$.

Parece obvio que el aumentar recursos (estados) suponga un aumento en la capacidad de los algoritmos (máquinas de Turing), por tanto es coherente la observación 2. Pero si consideramos todos los posibles usos que se pueden hacer de los nuevos recursos, habrá muchos algoritmos que usen los recursos para retrasar su tarea, y en promedio los nuevos algoritmos tardarán mucho más que aquellos que por tener menos recursos estaban más obligados a optimizarlos. Por tanto, comprendemos también la afirmación 3.

**Referencias**